\documentstyle[a4,12pt,epsfig,graphicx]{article}
\begin{document}
\titlepage
\begin{flushright}
CERN-TH/2000-329\\
DAMTP-2000-118\\
T/00-153\\
hep-th/0011045 \\
\end{flushright}
\vskip 1cm
\begin{center}
{ \Large
\bf Cosmological Solutions of Supergravity in Singular Spaces}
\end{center}

\vskip 1cm
\begin{center}
{\large Ph. Brax\footnote{email: philippe.brax@cern.ch} }
\end{center}
\vskip 0.5cm
\begin{center}
Theoretical Physics Division, CERN\\
CH-1211 Geneva 23\footnote { On leave of absence from  Service de Physique Th\'eorique, 
CEA-Saclay F-91191 Gif/Yvette Cedex, France}\\
\end{center}
\vskip .2 cm
\begin{center}
{\large A. C. Davis\footnote{email: a.c.davis@damtp.cam.ac.uk}}
\end{center}
\vskip .5 cm
\begin{center}
DAMTP, Centre for Mathematical Sciences, Cambridge University,
Wilberforce Road, Cambridge, CB3 0WA, UK. 
\end{center}

\vskip 2cm
\begin{center}
{\bf  Abstract}
\end{center}
\vskip .2 cm
\noindent
We study brane-world solutions of 
five-dimensional supergravity in singular spaces. We exhibit a self-tuned four-dimensional cosmological constant when five-dimensional supergravity is broken by an arbitrary tension on the brane-world.  The brane-world metric 
is of the FRW type corresponding to a cosmological constant $\Omega_{\Lambda}=
\frac{5}{7}$ and an equation of state $\omega=-\frac{5}{7}$ which are consistent
with experiment.

\vskip .3in \baselineskip10pt{

}
\bigskip
\vskip 1 cm

\noindent
\newpage
\baselineskip=1.5\baselineskip
\section{Introduction}

An outstanding problem in cosmology is understanding the origin of the
cosmological constant\cite{wein,wit}. This has been compounded by the recent evidence
of a small, non-zero value in type IA supernova data\cite{supernova}, 
suggesting that the dark energy density is of the same order of magnitude as 
the matter density.
There has been some progress towards understanding this fundamental problem.
On the one hand, there have been various proposals that it may
arise as `dark energy' due to the evolution of a scalar field\cite{quintessence}. These
quintessence models have met with some success. However, the origin of the
scalar field has yet to be addressed in such models. There has been an
alternative proposal, that the cosmological constant may arise from a rather
deeper understanding of space-time in higher dimensionsal theories
\cite{rubakov}. In particular, \cite{dimopoulos} have suggested that the 
cosmological constant in our
Universe could be induced by the properties of the `bulk' in a five 
dimensional model. Ref \cite{dimopoulos} showed that it was possible to
tune the bulk gravitational dynamics so that the contribution of the standard
model vacuum energy density was carried off the brane-world into the bulk.
Further progress was made in this direction by considering the approach of
a `self-tuning' domain wall\cite{grojean&hollowood} and developed in
\cite{grojean&binetruy}, which also pointed out short comings in the 
approach. Both\cite{grojean&hollowood} and\cite{grojean&binetruy} used
five dimensional Einstein equations coupled to a scalar field to cancel
the vacuum energy of the brane tension, and in both cases the scalar
field became singular in the bulk at a finite distance from the brane
world. 

Considerable progress has been made in understanding the origin of brane-worlds
by\cite{kallosh}, who considered five-dimensional supergravity in singular
spaces. Their formalism properly accounts for the boundary conditions on
our brane-world. Indeed, the conditions arising for a BPS solution are just
those arising from the junction conditions in\cite{grojean&binetruy}. As
a consequence, the natural framework in which to realise the possibility of
understanding the cosmological constant as arising from gravitational 
solutions in extra dimensions seem to be that of five dimensional supergravity
in singular spaces.

Here we develop the idea of the cosmological constant in our brane
world being induced by the behaviour of the bulk solution. We consider
five dimensional supergravity in singular spaces, taking careful account
of the boundary conditions on the brane. When supersymmetry is broken,
a time dependent scale factor is induced on the brane world. This gives
rise to an induced cosmological constant and equation of state on
the brane world. The induced cosmological constant is independent of
the supersymmetry breaking parameter. Using an exponential
superpotential in the bulk and the simplest one-dimensional example
of vector multiplet theory \cite{louis} there
are no free parameters. For a flat universe we obtain a cosmological
constant of $\Omega_\Lambda={5\over7}$ and corresponding matter density
$\Omega_m={2\over7}$, consistent
with experiment. In our equation of state we obtain $\omega=-{5\over7}$. 
Moreover, we show that the
form of our potential is the only one which gives consistent cosmology
on the brane world. Our solution has an evolving singularity in the bulk.
If the singularity hits the
brane world, then the five dimensional supergravity approximation used
here breaks down and the full string theory must be used.

In the next section we introduce  supergravity in singular spaces, considering
the structure of the bulk theory and deriving the special solutions which
we use in the following section. In section 3 we consider cosmological
solutions. Starting with static solutions, we show that time dependent
solutions arise when supersymmetry is broken on the brane.
The induced metric on the brane is of FRW type.
We then study the luminosity distance and deduce the acceleration parameter.
This allows us to infer the value of the effective cosmological constant
on the brane. The FRW dynamics appears to be due to a perfect fluid
whose equation of state always respects the dominant energy condition.
In the conclusions we comment on the possibility of connecting our analysis
with the usual matter dominated cosmological eras and the possibility
of brane-world quintessence.

\section{Supergravity in Singular Spaces}
In this section we shall recall the main ingredients of supergravity in 
singular spaces\cite{kallosh}.
The newly constructed supergravity theory differs from the
usual five-dimensional supergravity theories since space-time  boundaries
are taken into account. These boundaries provide new terms in the Lagrangian
and require new fields in order to close the supersymmetry algebra and
ensure the invariance of the Lagrangian.

More precisely space-time is supposed to be non-compact four dimensional
Minkowski space enlarged to five dimensions by the adjunction
of a $Z2$ orbifold of a circle.
The two boundaries of the interval are identified as branes, in particular
the brane at the origin is identified with our brane-world.
The bulk physics far from the boundaries is identical to
five dimensional supergravity coupled to $n-1$ vector multiplets. 
Let us recall briefly the structure of the bulk theory.

The vector multiplets comprise scalar fields $\phi^i$ parameterizing the manifold
\begin{equation}
C_{IJK}h^I(\phi) h^J(\phi) h^K(\phi)=1
\end{equation}
with the functions $h^I(\phi), I=1\dots n$ playing the role of auxiliary variables. In string theory the symmetric tensor $C_{IJK}$ has the meaning of an intersection tensor. Defining the metric 
\begin{equation}
G_{IJ}=-2C_{IJK}h^{K}+3h_Ih_J
\end{equation}
where $h_I=C_{IJK}h^Jh^K$, the bosonic part of the Lagrangian reads
\begin{equation}
S_{bulk}=\frac{1}{2\kappa_5^2}\int \sqrt {-g_5}(R-\frac{3}{4}(g_{ij}\partial_{\mu}
\phi^i\partial^{\mu}\phi^j+V))
\label{lag}
\end{equation}
where the induced metric is
\begin{equation}
g_{ij}=2 G_{IJ}\frac{\partial h^I}{\partial \phi^i}\frac{\partial h^J}{\partial \phi^j}
\end{equation}
and the potential is given by
\begin{equation}
V=W_iW^i-W^2.
\end{equation}
The superpotential $W$ defines the dynamics of the theory.
It is given  by 
\begin{equation}
W=4\sqrt\frac{2}{3}g h^IV_I
\end{equation}
where $g$ is a gauge coupling constant and the $V_I$'s are real number such that the
$U(1)$ gauge field is $A^I_{\mu}V_I$. The vectors   $A^I_{\mu}$ belong to the vector multiplets.  

The $Z2$ orbifold implies that a boundary action has to be incorporated.
The boundary action depends on two new fields. There is a supersymmetry singlet  $G$
and a four form $A_{\mu\nu\rho\sigma}$. One also introduces a modification of the bulk action by
replacing $g\to G$ and adding  a direct coupling
\begin{equation}
S_A=\frac{2}{4! \kappa_5^2}\int d^5x \epsilon^{\mu\nu\rho\sigma\tau}A_{\mu\nu\rho\sigma}\partial_{\tau}G.
\end{equation}
The boundary action is taken as
\begin{equation}
S_{bound}=-\frac{1}{\kappa_5^2}\int d^5x (\delta_{x_5}-\delta_{x_5-R})(\sqrt {-g_4}\frac{3}{2}W +\frac{2g}{4!}\epsilon^{\mu\nu\rho\sigma}A_{\mu\nu\rho\sigma}).
\end{equation}
The supersymmetry algebra closes on shell where
\begin{equation}
G(x)=g\epsilon (x_5),
\end{equation}
and $\epsilon (x_5)$ jumps from -1 to 1 at the origin of the fifth dimension.
On shell the bosonic Lagrangian reduces to  (\ref{lag}) coupled to the 
boundaries as,
\begin{equation}
S_{bound}=-\frac{3}{4\kappa_5^2}\int d^5x (\delta_{x_5}-\delta_{x_5-R})(\sqrt {-g_4}f),
\end{equation}
where we have defined
\begin{equation}
f=2W.
\label{rel}
\end{equation}
The equations of motion of such a 
Lagrangian show a BPS property and can be rewritten in a first order form.

Let us give the simplest example of models based on supergravity in singular spaces, i.e.
one vector multiplet scalar
such that the only component of the symmetric tensor $C_{IJK}$ is $C_{122}=1$.
The  moduli space of vector multiplets is then defined by the 
algebraic relation
\begin{equation}
h^1(h^2)^2=1.
\end{equation}
This allows to parameterize this  manifold using the coordinate 
$\phi$ such that 
 $h^1$ is proportional to $e^{\sqrt{\frac{1}{3}}\phi}$
and $h^2$ to $e^{-\phi/2\sqrt 3}$. The induced metric $g_{\phi\phi}$ can be
seen  to be one. The most general superpotential is a linear combination of the two exponentials
$W=ae^{\sqrt{\frac{1}{3}}\phi}+be^{-\phi/2\sqrt 3}$.
We will analyse this example in the next section.

\section{Cosmological Solutions}

In the following we shall restrict ourselves to a single scalar $\phi$ and normalize $g_{\phi\phi}=1$.
Let us look for static solutions with the metric
\begin{equation}
ds^2=e^{-A/2}dx_i^2+dx_5^2.
\end{equation}
The equations of motions  read
\begin{eqnarray}
A''&=&\phi'^2+f(\phi)\delta_{x_5}\nonumber \\
A'^2&=&\phi'^2 -V(\phi)\nonumber \\
\phi'' -A'\phi'&=&\frac{1}{2}\frac{\partial V}{\partial \phi}+
 \frac{\partial f}{\partial \phi}\delta_{x_5}\nonumber \\
\end{eqnarray}
leading to the junction conditions 
\begin{eqnarray}
\Delta(\frac{\partial W}{\partial \phi})&=&\frac{\partial f}{\partial \phi}\vert_{x_5=0}\nonumber \\
\Delta W&=& f\vert_{x_5=0}\nonumber \\
\end{eqnarray}
where the left-hand sides represent the discontinuity across the origin.
Notice that these conditions are automatically satisfied due
to the assignment (\ref{rel}).
The equations of motion can be written in BPS form
\begin{eqnarray}
\phi'&=&\frac{\partial W}{\partial \phi}\nonumber \\
A'&=&W.\nonumber \\
\end{eqnarray}
These equations depend explicitly on the choice of the superpotential $W$.

Here we will use the following model. We assume that there exists a field
$\phi$ with $g_{\phi\phi}=1$ such that one can choose the 
Fayet-Iliopoulos scalars $V_I$ 
in such a way that
\begin{equation}
W(\phi)=\xi e^{\alpha\phi}
\end{equation}
where $\xi$ is a characteristic scale.
Here the parameter $\alpha$ will play a crucial role. In the previous section 
we have shown that
$\alpha=\sqrt{\frac{1}{3}},\ -\sqrt{\frac{1}{12}}$ correspond to a 
one dimensional moduli space.

It is easy to show that the solutions to the BPS conditions are
\begin{eqnarray}
\phi&=&-\frac{1}{\alpha}\ln(1-\alpha^2\xi\vert x_5\vert)\nonumber\\
A&=&-\frac{1}{\alpha^2}\ln(1-\alpha^2\xi\vert x_5\vert).
\nonumber\\
\end{eqnarray}
Notice that we have chosen the scale factor 
\begin{equation}
a^2=e^{-A/2}
\end{equation}
to be normalized to one on the brane-world at $y=0$.
Another feature of this solution is the existence of a singularity at
\begin{equation}
x_{5*}= \frac{1}{\alpha^2\xi }.
\end{equation}
This implies that the supergravity description breaks down 
in the vicinity of $x_{5*}$, and one has to use the full string theory 
underlying this approach.

Since nature is not supersymmetric, we deform our model to introduce
supersymmetry breaking on the brane world  in a phenomenological
way. A viable theory should incorporate the standard model of particle physics
at low energy and describe the coupling between the standard model fields 
and the bulk fields living in five dimensions. We can modify the
previous study in a minimal way in order to take into account the
matter fields living on the brane. Let us assume that these matter fields
couple universally to the superpotential $W(\phi)$ 
\begin{equation}
-\frac{3}{2\kappa_5^2}\int d^4x \sqrt {-g_4}W(\phi)V(\Phi)
\end{equation}
and that to a good approximation the matter fields are fixed to 
their vev's -in particular we exclude time-dependent phenomena such as 
inflation-
in such a way that the effective coupling becomes 
\begin{equation}
f(\phi)=2TW(\phi)
\end{equation}
where 
\begin{equation}
T\equiv V(<\Phi>)
\end{equation}
encapsulates supersymmetry breaking for $T\ne 1$ which occurs only on the 
brane world. We assume that the bulk is still supersymmetric. In particular 
the tension $T$ is subject to radiative corrections and phase transitions. 
We will now discuss the deformations
of the previous static solutions when the supersymmetry breaking parameter 
$T$ is turned on.

We can  generate time-dependent conformally flat solutions from the static
solutions. This is most easily achieved by using a boost along
the $x_5$ direction.  To do so we shall first introduce conformal coordinates
so that the metric becomes
\begin{equation}
ds^2_5=a^2(u)(dx^2+du^2).
\end{equation}
Under a boost the new solutions of the bulk equations of motions are 
\begin{eqnarray}
a(u,\eta)&=& \sqrt{1-h^2} a(u+h\eta,\xi)\nonumber \\
\phi(u,\eta)&=& \phi(u+h\eta,\xi)\nonumber \\
\end{eqnarray}
where $x_1\equiv \eta$ is the conformal time.
The junction conditions for conformally flat metrics read\footnote{
The first of these conditions stems from the Israel conditions relating
the extrinsic curvature tensor $K_{ij}=\partial_n h_{ij}/2$, where $h_{ij}$
is the metric on the hypersurface orthogonal to the normal vector $\partial_n$,
and the energy momentum tensor on the brane world. In our case this reads
$\Delta (K_{ij}-Kh_{ij})=3fh_{ij}/4\vert_{\partial M}$ which leads to the first condition.}
\begin{eqnarray}
\Delta(\partial_n A)&=&f\vert_{\partial M}\nonumber \\
\Delta (\partial_n\phi)&=&\frac{\partial f}{\partial \phi}\vert_{\partial M}\nonumber \\
\end{eqnarray}
where the normal vector is $\partial_n\equiv a^{-1}\partial_u$ and $\partial M$ is the brane world.
In order to verify these conditions we will use a transformation 
which rescales  the conformal factor and the potential, i.e. the scale $\xi$
\begin{equation}
a\to \lambda a,\ V\to \frac{V}{\lambda^2}.
\end{equation}
The new solution of the bulk equations of motion is given by
\begin{eqnarray}
\tilde a(u,\eta)&=&  \frac{\sqrt{1-h^2}}{\lambda}a(u+h\eta,\frac{\xi}{\lambda})\nonumber \\
\tilde\phi(u,\eta)&=& \phi(u+h\eta,\frac{\xi}{\lambda} ).\nonumber \\
\label{sol}
\end{eqnarray}
One can  now use the BPS equations satisfied by $(a,\phi)$ to deduce that
\begin{eqnarray}
\partial_{\tilde n}\tilde \phi&=&\frac{1}{\sqrt{1-h^2}}\frac{\partial W}{\partial \tilde \phi}(\tilde\phi)\nonumber \\
\partial_{\tilde n}\tilde A&=& \frac{1}{\sqrt{1-h^2}}W (\tilde\phi)\nonumber \\
\end{eqnarray}
where $\tilde a= e^{-\tilde A/4}$ and the normal vector is 
\begin{equation}
\partial_{\tilde n}={\tilde a}^{-1}\partial_u.
\end{equation}
Under this rescaling the junction conditions become 
\begin{eqnarray}
\Delta(\frac{\partial  W}{\partial \tilde \phi})&=&T\sqrt{1-h^2}\frac{\partial f}
{\partial \tilde\phi}\vert_{\partial M}\nonumber \\
\Delta( W)&=&T\sqrt{1-h^2} f\vert_{\partial M}\nonumber\\
\end{eqnarray}
which are automatically satisfied thanks to the identity $ f=2  W$
provided that
\begin{equation}
h=\pm \frac{\sqrt{T^2-1}}{T}.
\end{equation}
Notice that this requires $T>1$. For smaller values of the breaking parameter 
the solutions are not defined and we thus need to go beyond the approximations
used here. 

In the following we choose $\lambda =\sqrt{1-h^2}$ for clarity.
The cosmological solutions are then 
\begin{equation}
a(u,\eta)=(\frac{u+h\eta}{u_0(h)})^{1/(4\alpha^2-1)}.
\end{equation}
The singularity is now located at 
\begin{equation}
u_*=-h\eta
\end{equation}
and the brane world at
\begin{equation}
u_0(h)\equiv \sqrt{1-h^2}{u_0}=\frac{\sqrt{1-h^2}}{(\frac{1}{4}-\alpha^2)\xi}.
\end{equation}
Depending on the sign of $h$ the singularity either converges
towards the brane-world or recedes towards the other end of the fifth dimension.
In either cases the supergravity approximation breaks down in finite time.
In particular the collision between the singularity and the brane-world
occurs at
\begin{equation}
\eta_0=-\frac{u_0(h)}{h}.
\end{equation}
Let us now focus on the brane-world.
The induced metric is  of the FRW type 
\begin{equation}
ds^2_{BW}=a^2(\eta)(-d \eta^2+ dx^idx_i)
\end{equation}
where  the scale factor is
\begin{equation}
a(\eta)=(1-\frac{\eta}{\eta_0})^{1/(4\alpha^2-1)}.
\end{equation}
The type of geometry on the brane-world depends on $\alpha$ and $h$.
Notice that there is always a singularity, either in the past or in the future.

The four dimensional Planck constant appears to be time-dependent
\begin{equation}
M^2_p=2M_5^3\int_{-h\eta }^{u_0(h)}\frac{a^{3}(u,\eta)}{a^2(\eta)}du.
\end{equation}
 This
leads to
\begin{equation}
M^2_p=\frac{4T}{2\alpha^2+1}\frac{M_5^3}{\xi} a(\eta)^{4\alpha^2}
\end{equation}
Now one can perform a Weyl rescaling $ds^2_4\to  T a(\eta)^{4\alpha^2} ds^2_4$
of the induced metric on the brane-world. This makes the Planck mass constant
\begin{equation} 
M^2_p=\frac{4}{2\alpha^2+1}\frac{M_5^3}{\xi}.
\end{equation}
The new scale factor becomes
\begin{equation}
a(\eta)=T (1-\frac{\eta}{\eta_0})^{(2\alpha^2+1)/(4\alpha^2-1)}.
\end{equation}
Notice that the exponent is now $(2\alpha^2+1)/(4\alpha^2-1)$ instead of $1/(4\alpha^2-1)$.
In particular if one uses cosmic time defined by $dt=a(\eta)d \eta$
there is a  singularity 
at $t_0=(1/3+1/6\alpha^2)T\eta_0$. 
More precisely the scale factor
becomes
\begin{equation}
a(t)=T(1-\frac{t}{t_0})^{1/3+1/6\alpha^2}
\end{equation} 
leading to the Hubble parameter
\begin{equation}
H=\frac{2\alpha^2+1}{6\alpha^2 (t-t_0)}.
\end{equation}
The universe is decelerating when $t_0>0$, i.e. when the singularity converges
to the brane-world.
It is accelerating when $t_0<0$ and the singularity recedes away from the brane-world. 
Notice that the supersymmetry breaking parameter appears in the overall normalization of the 
scale factor and in the time scale $t_0$. A variation of the parameter $T$ implies an
adaptation of the scale factor. However  the characteristic exponent is independent of $T$.

Let us now analyse the cosmological implications of these FRW models.
In particular it is phenomenologically highly relevant to
study the luminosity distance defined by
\begin{equation}
d_L=(1+z)\int_0^z\frac{dz'}{H(z')}
\end{equation}
where $z$ is the red-shift factor
\begin{equation}
\frac {a(t)}{a(0)}=\frac{1}{1+z}.
\end{equation}
The small $z$ expansion reads in general
\begin{equation}
d_L\sim \frac{1}{H_0}(z+ \frac{1-q_0}{2}z^2+o(z^3))
\end{equation}
and the acceleration parameter is directly related to the
matter energy density $\Omega_m$ and the effective cosmological constant
density $\Omega_{\Lambda}$
\begin{equation}
\Omega_{\Lambda}=\frac{\Omega_m}{2}-q_0.
\end{equation}

Notice that from the observer's point of view a finite and
small cosmological constant, i.e. of the order of the critical density, 
appears as a consequence of the interpretation of the four dimensional
FRW geometry as resulting from the dynamics in five dimensions. In our
 models the effective cosmological constant becomes
\begin{equation}
\Omega_{\Lambda}=\frac{\Omega_m}{2}+1-\frac{6\alpha^2}{1+2\alpha^2}.
\end{equation}
This can accommodate a small and positive cosmological constant, which is 
independent of the supersymmetry breaking parameter.

Another relevant observable from the four dimensional point of view is
the equation of state
of the brane-world $p=\omega \rho$ relating the pressure to the energy density. It is given by
\begin{equation}
\omega=-1+\frac{4\alpha^2}{1+2\alpha^2}.
\end{equation}
It can take any value between -1 and 1, i.e. it never violates the
dominant energy condition. The observer will therefore be driven to
conclude that the four dimensional dynamics of the universe is due
to some matter with the previous equation of state.

The previous model has been obtained for a particular choice of superpotential.
However it is easy to calculate the acceleration parameter 
before the Weyl rescaling. It happens to be $-1+4(\frac{d\ln W}{d\phi})^2\vert_{\phi_0}$ where $\phi_0$ is the value of $\phi$ on the brane. Hence  only an exponential superpotential yields a boundary value independent
result for the acceleration parameter.

In the case of the one-dimensional moduli space  discussed earlier with 
$\alpha=\sqrt {\frac{1}{3}}$, the 
induced cosmological constant is negative.
In the other case of $\alpha= -\sqrt{\frac{1}{12}}$, the cosmological 
constant and the equation of state are  given by
\begin{equation}
\Omega_{\Lambda}=\frac{\Omega_m}{2}+\frac{4}{7},\ \omega =-\frac{5}{7}.
\end{equation}
Imposing a flat universe, $\Omega_0\equiv \Omega_m+\Omega_{\Lambda}=1$, yields 
\begin{equation}
\Omega_{\Lambda}=\frac{5}{7},\ \Omega_m=\frac{2}{7},\ \omega =-\frac{5}{7}.
\end{equation}
These numbers are compatible with the current experimental results from type
Ia supernovae and CMB anisotropies. 

In brief we have exhibited a mechanism where the effective cosmological
constant is naturally of the right order of magnitude. This springs from the
five dimensional origin of the FRW dynamics of the universe. Moreover we have
shown that including the non-supersymmetric effects of the matter fields
on the brane world does not affect the FRW features of the universe, in 
particular
the cosmological constant is not modified by variations of the potential energy
of the matter fields.

\section{Conclusions}
In this letter we have realised the original suggestion of \cite{rubakov}
and \cite{dimopoulos} that the observed cosmological constant is induced
on our brane world by the dynamics of fields in extra dimensions. In our
case we exhibited a self-tuned cosmological constant induced by five 
dimensional supergravity once supersymmetry was broken on the brane world.
Further, our mechanism gave a relationship between the matter density and
the dark energy density. Imposing a flat universe resulted in values
consistent with observation. Our induced cosmological constant is
independent of the value of the supersymmetry breaking parameter. 

Our mechanism relies on supersymmetry breaking on the brane world
inducing time dependent solutions, which are of the FRW type. The
FRW dynamics are of the perfect fluid type, with an equation of
state that respects the dominant energy condition and is consistent
with experiment. 
 
At present we have considered static solutions in the bulk with
cosmological solutions being induced on the brane world once supersymmetry
is broken. This has given us an accelerating universe. In order to explore
the transition between an accelerating universe and matter domination we
need to consider the bulk solution in a perfect fluid. For example, 
\cite{kennedy} has considered a self-tuned domain wall in five dimensional
gravity coupled to a scalar field with bulk fluid, obtaining matter 
dominated FRW dynamics induced on the brane. Coupling this to our 
mechanism for an accelerating universe should result in a rich cosmology,
such as a modified Friedmann equation on the brane-world.
This is in progress.

\section{Acknowledgements}
We wish to thank C. Grojean and N. Turok for discussions. 
This work is supported in part by  PPARC (A.C. D.). 

\def\Journal#1#2#3#4{{#1}{\bf #2}, #3 (#4)}
\def\NPB{Nucl.\ Phys.\ {\bf B}}
\def\PLB{Phys.\ Lett.\ {\bf B}}
\def\PRL{Phys.\ Rev.\ Lett.\ }
\def\PRD{Phys.\ Rev.\ D }
\def\JPA{J.\ Phys.\ {\bf A}}

\end{document}